\documentclass[aps,onecolumn,showpacs,showkeys,nofootinbib]{revtex4}
\usepackage{epsfig}
\usepackage{amsmath}
\usepackage{amsfonts}
\usepackage{amssymb}
\usepackage{graphicx}
\usepackage{colordvi}
\begin{document}

\title{Hydrostatic equilibrium  and stellar structure in  $f(R)$-gravity}

\author{S. Capozziello$^1$\footnote{e-mail address: capozziello@na.infn.it}, M. De Laurentis$^1$\footnote{e-mail address: felicia@na.infn.it}, S.D. Odintsov$^2$\footnote{e-mail address: odintsov@aliga.ieec.uab.es. Also at TSPU, Tomsk, Russia.}, A. Stabile$^3$\footnote{e-mail address: arturo.stabile@gmail.com}}

\affiliation{$^1$Dipartimento di Scienze Fisiche, Universita'
di Napoli {}``Federico II'', INFN Sez. di Napoli, Compl. Univ. di
Monte S. Angelo, Edificio G, Via Cinthia, I-80126, Napoli, Italy;\\
$^2$Institucio Catalana de Recerca i Estudis Avancats (ICREA) and Institut de Ciencies de l Espai (IEEC-CSIC),
Campus UAB, Facultat de Ciencies, Torre C5-Par-2a pl, E-08193 Bellaterra (Barcelona), Spain;\\
$^3$Dipartimento di Ingegneria,
Universita' del Sannio, Palazzo Dell'Aquila Bosco Lucarelli, Corso Garibaldi, 107 - 82100, Benevento, Italy.}

\begin{abstract}

We investigate the  hydrostatic equilibrium of stellar structure by taking into account the modified Lan\'{e}-Emden equation coming out from $f(R)$-gravity. Such an equation  is   obtained  in metric approach by considering the Newtonian limit of $f(R)$-gravity, which gives rise to a modified Poisson equation, and then  introducing a relation between  pressure and density with polytropic index $n$.  The  modified equation results  an integro-differential equation, which, in the limit $f(R)\,\rightarrow\,R$, becomes the standard Lan\'{e}-Emden equation. We find the radial profiles of gravitational potential by solving for some values of $n$.  The comparison of solutions with those coming from General Relativity shows that they are compatible and physically relevant.
\end{abstract}
\date{\today}
\maketitle

\section{Introduction}
Extended Theories of Gravity (ETG) \cite{book} are  a new paradigm of modern physics aimed to address several shortcomings coming out in the study of gravitational interaction at ultra-violet and infra-red scales. In particular,  instead of  introducing  unknown fluids, the approach consists in extending  General Relativity (GR) by taking into account   generic  functions of curvature invariants. These functions can be physically motivated and  capable of addressing phenomenology at galactic, extragalactic, and cosmological scales \cite{review}.

 This viewpoint does not require to find out
candidates for dark energy and dark matter at fundamental level
(not  detected up to now),  but takes into account
only the observed ingredients (\emph{i.e.} gravity, radiation and
baryonic matter), changing the \emph{l.h.s.} of the field equations. Despite of this modification, it is in
agreement with the spirit of GR since the only
request is that the Hilbert-Einstein action should be generalized
asking for a gravitational interaction  acting, in principle,  in
different ways at different scales but preserving the robust results of GR at local and Solar System scales (see \cite{book} for a detailed discussion). 
This is the case of $f(R)$-gravity which reduces to GR as soon as $f(R)\,\rightarrow\,R$.

Other issues as, for example, the observed Pioneer anomaly problem
\cite{anderson} can be framed into the same approach
\cite{bertolami} and then, apart the cosmological dynamics, a
systematic analysis of such theories urges at short scales and in
the low energy limit.

On the other hand, the strong gravity regime \cite{psaltis}  is
another way to check the viability of these theories. In general the formation and the evolution of stars can be considered suitable test-beds for
Alternative Theories of Gravity.  Considering the case of $f(R)$-gravity,  divergences
stemming from the functional form of $f(R)$ may prevent the existence of
relativistic stars in these theories \cite{briscese}, but thanks
to the chameleon mechanism, introduced by   Khoury and  Weltman \cite{weltman}, the possible problems jeopardizing the existence
of these objects may be avoided \cite{tsu}. Furthermore, there are
also numerical solutions corresponding to static star configurations with
 strong gravitational fields \cite{babi} where the choice of the
equation of state  is crucial for the existence of solutions.

 Furthermore some observed stellar systems  are incompatible with the
standard models of stellar structure. We refer to anomalous neutron stars, the so called 
"magnetars"  \cite{mag}  with masses larger than their expected  Volkoff mass.
It  seems that, on particular length scales, the gravitational force is larger
or smaller than the corresponding GR value.
For example, a modification of the Hilbert-Einstein Lagrangian, consisting of  $R^2$ terms,
enables a major attraction while a $R_{\alpha\beta}R^{\alpha\beta}$ term gives a repulsive  
contribution \cite{stabile_2}. Understanding 
on which scales the modifications to GR are working or what is the weight of corrections 
to gravitational potential is a crucial point that could confirm or rule out these extended approaches to gravitational interaction.

The plan of paper is the following: In Sec. \ref{classic}, we review briefly the classical hydrostatic problem for stellar structures. In Sec. \ref{FEnewtonian} we derive the  
Newtonian limit of $f(R)$-gravity obtaining the modified Poisson equation. The modified Lan\'{e}-Emden equation is obtained in Sec. \ref{LEsection}  and its structure is compared  with respect to the standard one. In Sec.\ref{solutions}, we show the analytical solutions of standard Lan\'{e}-Emden equation and compare them with those obtained perturbatively  from $f(R)$-gravity.  With help of plot we can compare between them all results. Discussion and conclusions are drawn in Sec. \ref{conclu} .

\section{Hydrostatic equilibrium of stellar structures}
\label{classic}

The  condition of hydrostatic equilibrium for  stellar structures  in Newtonian dynamics is achieved by   considering the  equation

\begin{equation}\label{19.1}
\frac{dp}{dr}\,=\,\frac{d\Phi}{dr}\rho\,,
\end{equation}
where $p$ is the pressure, $-\Phi$ is the gravitational potential, and $\rho$ is the density \cite{kippe}.   Together with  the above equation, the Poisson equation

\begin{equation}\label{19.2}
\frac{1}{r^2}\frac{d}{dr}\left(r^2\frac{d\Phi}{dr}\right)\,=\,-4\pi G\rho\,,
\end{equation}
gives the gravitational potential as  solution for a given matter density $\rho$. 
Since we are taking into account only static and stationary situations, here  we consider only time-independent solutions \footnote{The radius $r$  is assumed as the spatial coordinate. It  varies from $r\,=\,0$ at the center to  $r\,=\,\xi$ at the surface of the star}.
In general, the temperature $\tau$ appears in Eqs. (\ref{19.1}) and (\ref{19.2})  the density satisfies an equation of state of the form $\rho\,=\,\rho(p,\tau)$. In any case, we assume that there exists a polytropic relation between $p$ and $\rho$ of the form

\begin{equation}\label{19.3}
p\,=\,K\rho^\gamma\,,
\end{equation}
where $K$ and $\gamma$ are constant. Note that  $\Phi\,>\,0$ in the interior of the model  since we define the gravitational potential as $-\Phi$. The polytropic constant $K$ is fixed and can be obtained as a combination of fundamental  constants. However there are several realistic  cases where $K$ is not fixed and another equation for its evolution is needed.  The constant $\gamma$  is the {\it polytropic exponent }. Inserting  the polytropic equation of state into  Eq. (\ref{19.1}), we obtain
\begin{equation}\label{19.6}
\frac{d\Phi}{dr}\,=\,\gamma K \rho^{\gamma-2}\frac{d\rho}{dr}\,.
\end{equation}
For  $\gamma\neq1$,  the above equation can be integrated giving

\begin{eqnarray}\label{densita}
\frac{\gamma K}{\gamma-1}\rho^{\gamma-1}\,=\,\Phi\,\,\,\,\rightarrow\,\,\,\,\rho\,=\,\biggl[\frac{\gamma-1}{\gamma K}\biggr]^{\frac{1}{\gamma-1}}\Phi^{\frac{1}{\gamma-1}}\,\doteq\,A_n\Phi^n
\end{eqnarray}
where we have chosen the integration constant to give $\Phi=0$ at surface $(\rho=0$). The constant $n$ is called the {\it polytropic index} and is defined as  $n=\frac{1}{\gamma-1}$. Inserting the relation (\ref{densita})  into the Poisson equation, we obtain a differential equation for the gravitational potential
\begin{equation}\label{19.8}
\frac{d^2\Phi}{dr^2}+\frac{2}{r}\frac{d\Phi}{dr}\,=\,-4\pi G A_n\Phi^n\,.
\end{equation}
Let us define now the dimensionless variables

\begin{eqnarray}\label{trans1}
\left\{\begin{array}{ll}
z\,=\,|\mathbf{x}|\sqrt{\frac{\mathcal{X}A_n\Phi_c^{n-1}}{2}}\\\\
w(z)\,=\,\frac{\Phi}{\Phi_c}\,=\,(\frac{\rho}{\rho_c})^\frac{1}{n}
\end{array}\right.
\end{eqnarray}
where the subscript $c$ refers to the center of the star and  the relation between $\rho$ and $\Phi$ is given by Eq. (\ref{densita}). At the center  $(r\,=\,0)$,  we have $z\,=\,0$, $\Phi\,=\,\Phi_c$, $\rho\,=\,\rho_c$ and therefore $w\,=\,1$. Then 
Eq. (\ref{19.8}) can be written

\begin{eqnarray}\label{LE}
\frac{d^2w}{dz^2}+\frac{2}{z}\frac{dw}{dz}+w^n\,=\,0
\end{eqnarray}
This is the standard {\it Lane-Embden equation} describing the hydrostatic equilibrium of stellar structures in the Newtonian theory \cite{kippe}.
\section{The Newtonian limit of  $f(R)$-gravity}\label{FEnewtonian}
Let us start with a general class of  Extended Theories of Gravity given
by the action

\begin{eqnarray}\label{FOGaction}
\mathcal{A}\,=\,\int d^{4}x\sqrt{-g}[f(R)+\mathcal{X}\mathcal{L}_m]
\end{eqnarray}
where $f(R)$ is an analytic  function of the curvature invariant $R$.
 $\mathcal{L}_m$ is the minimally coupled ordinary matter Lagrangian density.
 In the metric approach, the field equations are
obtained by varying the action (\ref{FOGaction}) with respect to
$g_{\mu\nu}$. We get

\begin{eqnarray}\label{fieldequationFOG}
\left\{\begin{array}{ll}
f'R_{\mu\nu}-\frac{f}{2}g_{\mu\nu}-f_{;\mu\nu}+g_{\mu\nu}\Box f'=\mathcal{X}\,T_{\mu\nu}\\
\\
3\Box
f'+f'R-2f\,=\,\mathcal{X}\,T
\end{array}\right.
\end{eqnarray}

where the second equation is the trace of the field equations. Here,
${\displaystyle T_{\mu\nu}\,=\,\frac{-1}{\sqrt{-g}}\frac{\delta(\sqrt{-g}\mathcal{L}_m)}{\delta
g^{\mu\nu}}}$ is the the energy-momentum tensor of matter;
$T\,=\,T^{\sigma}_{\,\,\,\,\,\sigma}$ is the trace;
$f'\,=\,\frac{df(R)}{dR}$, $\Box\,=\,{{}_{;\sigma}}^{;\sigma}$ the d'Alembert operator and
$\mathcal{X}\,=\,8\pi G$.  We assume
$c\,=\,1$ is adopted. The conventions for Ricci's tensor are
$R_{\mu\nu}\,=\,{R^\sigma}_{\mu\sigma\nu}$; the Riemann
tensor is
${R^\alpha}_{\beta\mu\nu}\,=\,\Gamma^\alpha_{\beta\nu,\mu}+...$. The
affine connections  are the  Christoffel's symbols of the metric
$\Gamma^\mu_{\alpha\beta}\,=\,\frac{1}{2}g^{\mu\sigma}(g_{\alpha\sigma,\beta}+g_{\beta\sigma,\alpha}
-g_{\alpha\beta,\sigma})$. The  signature is $(+---)$.

In order to achieve the Newtonian limit of the theory
 the metric tensor $g_{\mu\nu}$  have to be approximated as follows 

\begin{eqnarray}\label{metric_tensor_PPN}
  g_{\mu\nu}\,\sim\,\begin{pmatrix}
  1-2\,\Phi(t,\mathbf{x})+\mathcal{O}(4)& \mathcal{O}(3) \\
  \\
  \mathcal{O}(3)& -\delta_{ij}+\mathcal{O}(2)\end{pmatrix}
\end{eqnarray}
where $\mathcal{O}(n)$ (with $n\,=$ integer) denotes the order of the expansion (see \cite{mio} for details).
The set of coordinates\footnote{The Greek index runs between $0$
and $3$; the Latin index between $1$ and $3$.} adopted is
$x^\mu\,=\,(t,x^1,x^2,x^3)$. The Ricci scalar formally becomes

\begin{eqnarray}
R\,\sim\,R^{(2)}(t,\mathbf{x})+\mathcal{O}(4)
\end{eqnarray}
The $n$-th derivative of Ricci function can be developed as

\begin{eqnarray}
f^{n}(R)\,\sim\,f^{n}(R^{(2)}+\mathcal{O}(4))\,\sim\,f^{n}(0)+f^{n+1}(0)R^{(2)}+\mathcal{O}(4)
\end{eqnarray}
here $R^{(n)}$ denotes a quantity of order  $\mathcal{O}(n)$.
From lowest order of field Eqs. (\ref{fieldequationFOG}), we
have $f(0)\,=\,0$ which trivially follows from the above assumption (\ref{metric_tensor_PPN}) for the metric.  This means that  the space-time
is asymptotically Minkowskian and we are discarding a cosmological constant term in this analysis\footnote{This assumption is quite natural since the contribution of a cosmological constant term is irrelevant at stellar level. }. Eqs. (\ref{fieldequationFOG})  at $\mathcal{O}(2)$-order, that is at Newtonian level, are

\begin{eqnarray}\label{PPN-field-equation-general-theory-fR-O2}
\left\{\begin{array}{ll}
R^{(2)}_{tt}-\frac{R^{(2)}}{2}-f''(0)\triangle
R^{(2)}\,=\,\mathcal{X}\,T^{(0)}_{tt}
\\\\
-3f''(0)\triangle
R^{(2)}-R^{(2)}\,=\,\mathcal{X}\,T^{(0)}
\end{array}\right.
\end{eqnarray}
where $\triangle$ is the Laplacian in the flat space, $R^{(2)}_{tt}\,=\,-\triangle\Phi(t,\mathbf{x})$ and, for the sake of  simplicity, we set $f'(0)\,=\,1$. We recall that the energy-momentum tensor  for a perfect fluid  is

\begin{eqnarray}
T_{\mu\nu}\,=\,(\epsilon+p)\,u_{\mu} u_{\nu}-p\,g_{\mu\nu}
\end{eqnarray}
where $p$ is the pressure  and $\epsilon$ is the energy density. Being the pressure contribution negligible in the field equations in Newtonian approximation,  we have

\begin{eqnarray}\label{HOEQ}
\left\{\begin{array}{ll}
\triangle\Phi+\frac{R^{(2)}}{2}+f''(0)\triangle R^{(2)}\,= \,-\mathcal{X}\rho
\\\\
3f''(0)\triangle R^{(2)}+R^{(2)}\,=\,-\mathcal{X}\rho
\end{array}\right.
\end{eqnarray}
where $\rho$ is now the mass density\footnote{Generally it is $\epsilon\,=\,\rho\,c^2$.}. We note that  for $f''(0)\,=\,0$ we have the  standard Poisson equation: $\triangle\Phi\,=\,-4\pi G\rho$. This means that as soon as the second derivative of $f(R)$ is different from zero, deviations from the Newtonian limit of GR emerge.

The gravitational potential $-\Phi$, solution of Eqs. (\ref{HOEQ}), has in general
a Yukawa-like behavior depending on a
characteristic length on which it evolves \cite{mio}. Then as it is evident
the Gauss theorem is not valid\footnote{It is worth noticing that also if the Gauss theorem does not hold, the Bianchi identities are always valid so the conservation laws are guaranteed. } since the force
law is not $\propto|\mathbf{x}|^{-2}$. The equivalence between a
spherically symmetric distribution and point-like distribution is
not valid and how the matter is distributed in the space is very
important \cite{mio,stabile,cqg}.

Besides the Birkhoff theorem results modified at Newtonian level:
the solution can be only factorized by a space-depending function
and an arbitrary time-depending function \cite{mio}. Furthermore
the correction to the gravitational potential is depending on
the only first two derivatives of $f(R)$ in $R\,=\,0$. This means that different analytical 
theories, from the third derivative perturbation terms on,  admit the same Newtonian
limit  \cite{mio,stabile}.

 Eqs. (\ref{HOEQ}) can be considered the \emph{modified Poisson equation}
for $f(R)$-gravity. They  do not depend on gauge condition choice \cite{cqg}.

\section{Stellar Hydrostatic Equilibrium in $f(R)$-gravity}\label{LEsection}

From the Bianchi identity,  satisfied by the field Eqs. (\ref{fieldequationFOG})),  we have

\begin{eqnarray}\label{equidr}
{T^{\mu\nu}}_{;\mu}\,=\,0\,\,\,\,\rightarrow\,\,\,\,\frac{\partial p}{\partial x^k}\,=\,-\frac{1}{2}(p+\epsilon)\frac{\partial \ln g_{tt}}{\partial x^k}
\end{eqnarray}
If the dependence on the temperature  $\tau$ is negligible, \emph{i.e.} $\rho\,=\,\rho(p)$,  this relation can be introduced into Eqs. (\ref{HOEQ}), which become a system of three equations for $p$, $\Phi$ and $R^{(2)}$ and can be solved without the other structure equations.

Let us suppose that matter satisfies still a polytropic equation $p\,=\,K\,\rho^\gamma$. If we introduce Eq.(\ref{densita}) into Eqs. (\ref{HOEQ}) we obtain an integro-differential equation for the gravitational potential $-\Phi$, that is 

\begin{eqnarray}\label{deltafi}
\triangle\Phi(\mathbf{x})+\frac{2\mathcal{X}A_n}{3}\Phi(\mathbf{x})^n\,=\,-\frac{m^2\mathcal{X}A_n}{6} \int d^3\mathbf{x}'\mathcal{G}(\mathbf{x},\mathbf{x}')\Phi(\mathbf{x}')^n
\end{eqnarray}
where ${\displaystyle \mathcal{G}(\mathbf{x},\mathbf{x}')\,=\,-\frac{1}{4\pi}\frac{e^{-m|\mathbf{x}-\mathbf{x}'|}}{|\mathbf{x}-\mathbf{x}'|}}$
is the Green function of  the  field operator $\triangle_\mathbf{x}-m^2$ for systems with spherical symmetry and ${\displaystyle m^2\,=\,-\frac{1}{3f''(0)}}$ (for details see \cite{stabile,cqg}). The integro-differential nature of Eq.(\ref{deltafi}) is the proof of the non-viability of Gauss theorem for $f(R)$-gravity. Adopting again the dimensionless variables

\begin{eqnarray}\label{trans}
\left\{\begin{array}{ll}
z\,=\,\frac{|\mathbf{x}|}{\xi_0}\\\\
w(z)\,=\,\frac{\Phi}{\Phi_c}
\end{array}\right.
\end{eqnarray}
where

\begin{eqnarray}\label{charpar}
\xi_0\,\doteq\,\sqrt{\frac{3}{2\mathcal{X}A_n\Phi_c^{n-1}}}
\end{eqnarray}
is a characteristic length linked to stellar radius $\xi$, Eq. (\ref{deltafi}) becomes

\begin{eqnarray}\label{LEmod}
\frac{d^2w(z)}{dz^2}+\frac{2}{z}\frac{d w(z)}{dz}+w(z)^n\,=\,\frac{m\xi_0}{8}\frac{1}{z}\int_0^{\xi/\xi_0}
dz'\,z'\,\biggl\{e^{-m\xi_0|z-z'|}-e^{-m\xi_0|z+z'|}\biggr\}\,w(z')^n
\end{eqnarray}
which is the \emph{modified Lan\'{e}-Emden equation} deduced from $f(R)$-gravity. Clearly  the particular $f(R)$-model is specified by the parameters $m$ and $\xi_0$. 
If $m\,\rightarrow\,\infty$ (\emph{i.e.} $f(R)\,\rightarrow\,R$),  Eq. (\ref{LEmod}) becomes Eq. (\ref{LE}). We are only interested in solutions of Eq. (\ref{LEmod}) that are finite at the center, that is for $z\,=\,0$. Since the center must be an equilibrium point,  the gravitational acceleration $|\mathbf{g}|\,=\,-d\Phi/dr\,\propto\,dw/dz$ must vanish for $w'(0)\,=\,0$. Let us assume we have solutions $w(z)$ of Eq.(\ref{LEmod}) that fulfill the  boundary conditions $w(0)\,=\,1$ and $w(\xi/\xi_0)\,=\,0$; then according to the choice (\ref{trans}), the radial distribution of  density is given by

\begin{eqnarray}
\rho(|\mathbf{x}|)\,=\,\rho_cw^n\,,\,\,\,\,\,\,\,\,\rho_c\,=\,A_n{\Phi_c}^n
\end{eqnarray}
and the pressure by

\begin{eqnarray}
p(|\mathbf{x}|)\,=\,p_cw^{n+1}\,,\,\,\,\,\,\,\,\,p_c\,=\,K{\rho_c}^\gamma
\end{eqnarray}

For $\gamma\,=\,1$ (or $n\,=\,\infty$) the integro-differential Eq. (\ref{LEmod}) is not correct. This means that the theory does not contain the case of isothermal sphere of ideal gas. 
In this case, the polytropic relation is $p\,=\,K\,\rho$. Putting this relation into Eq.(\ref{equidr}) we have
\begin{eqnarray}\label{densitaiso}
\frac{\Phi}{K}\,=\,\ln\rho-\ln\rho_c\,\,\,\,\rightarrow\,\,\,\,\rho\,=\,\rho_c\,e^{\Phi/K}
\end{eqnarray}
where  the constant of integration is chosen in such a way that the gravitational potential is zero at the center. If we introduce Eq.(\ref{densitaiso}) into Eqs. (\ref{HOEQ}), we have

\begin{eqnarray}\label{deltafiiso}
\triangle\Phi(\mathbf{x})+\frac{2\mathcal{X}\rho_c}{3}e^{\Phi(\mathbf{x})/K}\,=\,-\frac{m^2\mathcal{X}\rho_c}{6} \int d^3\mathbf{x}'\mathcal{G}(\mathbf{x},\mathbf{x}')e^{\Phi(\mathbf{x}')/K}
\end{eqnarray}
Assuming the dimensionless variables $z\,=\,\frac{|\mathbf{x}|}{\xi_1}$ and $w(z)\,=\,\frac{\Phi}{K}$ where $\xi_1\,\doteq\,\sqrt{\frac{3K}{2\mathcal{X}\rho_c}}$, Eq. (\ref{deltafiiso}) becomes

\begin{eqnarray}\label{LEmodiso}
\frac{d^2w(z)}{dz^2}+\frac{2}{z}\frac{d w(z)}{dz}+e^{w(z)}\,=\,\frac{m\xi_1}{8}\frac{1}{z}\int_0^{\xi/\xi_1}
dz'\,z'\,\biggl\{e^{-m\xi_1|z-z'|}-e^{-m\xi_1|z+z'|}\biggr\}\,e^{w(z')}
\end{eqnarray}
which is the \emph{modified "isothermal" Lan\'{e}-Emden equation} derived $f(R)$-gravity.

\section{ Solutions of the standard and modified   Lan\'{e}-Emden Equations}\label{solutions}
\begin{figure}[htbp]
  \centering
  \includegraphics[scale=1]{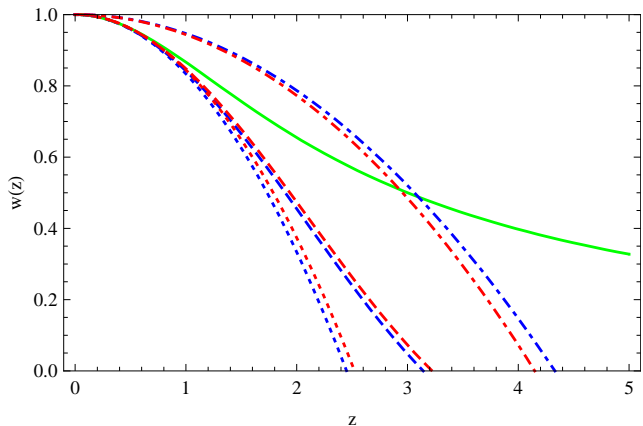}\\
  \caption{Plot of solutions (blue lines) of standard Lan\'{e}-Emden Eq. (\ref{LE}): $w^{(0)}_{GR}(z)$ (dotted line) and $w^{(1)}_{GR}(z)$ (dashed line). The green line corresponds to $w^{(5)}_{GR}(z)$. The red lines are the solutions of modified Lan\'{e}-Emden Eq. (\ref{LEmod}): $w^{(0)}_{_{f(R)}}(z)$ (dotted line) and $w^{(1)}_{_{f(R)}}(z)$ (dashed line). The blue dashed-dotted line is the  potential  derived from GR ($w_{GR}(z)$) and the red dashed-dotted line the  potential derived from  $f(R)$-gravity ($w_{_{f(R)}}(z)$) for a uniform spherically symmetric mass distribution. The assumed values are $m\xi\,=\,1$ and $m\xi_0\,=\,.4$. From a rapid inspection of these plots, the differences between GR and $f(R)$ gravitational potentials are clear and the tendency is that at larger radius $z$ they become more evident. }
  \label{fig}
\end{figure}

The task is now to solve the modified Lan\'{e}-Emden equation and compare its solutions to those of standard Newtonian theory.
Only for three values of $n$,  the solutions of Eq.(\ref{LE}) have analytical expressions \cite{kippe}

\begin{eqnarray}\label{LEsol}
&&n\,=\,0\,\,\,\,\rightarrow\,\,\,\,w^{(0)}_{GR}(z)\,=\,1-\frac{z^2}{6}\nonumber\\
&&n\,=\,1\,\,\,\,\rightarrow\,\,\,\,w^{(1)}_{GR}(z)\,=\,\frac{\sin z}{z}\\
&&n\,=\,5\,\,\,\,\rightarrow\,\,\,\,w^{(5)}_{GR}(z)\,=\,\frac{1}{\sqrt{1+\frac{z^2}{3}}}\nonumber
\end{eqnarray}
We label these solution with $_{GR}$ since they agree with the Newtonian limit of GR.
The surface of the polytrope of index $n$ is defined by the value $z\,=\,z^{(n)}$, where $\rho\,=\,0$ and thus $w\,=\,0$. For $n\,=\,0$ and $n\,=\,1$ the surface is reached for a finite value of $z^{(n)}$. The case $n\,=\,5$ yields a model of infinite radius. It can be shown that for $n\,<\,5$ the radius of polytropic models is finite; for $n\,>\,5$ they have infinite radius. From Eqs.(\ref{LEsol}) one finds $z^{(0)}_{GR}\,=\,\sqrt{6}$, $z^{(1)}_{GR}\,=\,\pi$ and $z^{(5)}_{GR}\,=\,\infty$. A general property of the solutions is that $z^{(n)}$ grows monotonically with the polytropic index $n$. In Fig. \ref{fig} we show the behavior of solutions $w^{(n)}_{GR}$ for $n\,=\,0,\,1,\,5$.
Apart from the three cases where analytic solutions are known, the classical Lan\'{e}-Emden Eq. (\ref{LE}) has to be be solved numerically, considering with the expression

\begin{eqnarray}\label{taydev}
w^{(n)}_{GR}(z)\,=\,\sum_{i\,=\,0}^\infty a^{(n)}_iz^i
\end{eqnarray}
for the neighborhood of the center. Inserting Eq.(\ref{taydev}) into Eq. (\ref{LE}) and by comparing coefficients one finds, at lowest orders, a classification of solutions by the index $n$, that is

\begin{eqnarray}\label{gensolGR}
w^{(n)}_{GR}(z)\,=\,1-\frac{z^2}{6}+\frac{n}{120}z^4+\dots
\end{eqnarray}
The case  $\gamma\,=\,5/3$ and $n\,=\,3/2$ is the non-relativistic limit  while the case  $\gamma\,=\,4/3$ and $n\,=\,3$ is the relativistic limit of a completely degenerate gas.

Also for modified Lan\'{e}-Emden Eq. (\ref{LEmod}),  we have an  exact solution for $n\,=\,0$. In fact, it is straightforward to find out 

\begin{eqnarray}\label{LEmodsol0}
w^{(0)}_{_{f(R)}}(z)\,=\,1-\frac{z^2}{8}+\frac{(1+m\xi)e^{-m\xi}}{4m^2{\xi_0}^2}\biggl[1-\frac{\sinh m\xi_0 z}{m\xi_0 z}\biggr]
\end{eqnarray}
where  the boundary conditions $w(0)\,=\,1$ and $w'(0)\,=\,0$ are satisfied. A comment on the GR limit (that is  $f(R)\rightarrow R$) of solution (\ref{LEmodsol0}) is necessary. In fact when we perform the limit $m\,\rightarrow\,\infty$, we do not recover  exactly $w^{(0)}_{GR}(z)$. The difference is in the definition of quantity $\xi_0$. In $f(R)$-gravity we have the definition (\ref{charpar}) while in GR it is ${\displaystyle \xi_0\,=\,\sqrt{\frac{2}{\mathcal{X}A_n\Phi_c^{n-1}}}}$, since in the first equation of (\ref{HOEQ}), when we perform $f(R)\rightarrow R$, we have to eliminate the trace equation condition. In general, this means that the Newtonian limit and the Eddington parameterization of different relativistic theories of gravity cannot coincide with those of GR (see \cite{eddington} for further details on this point).

The point $z_{_{f(R)}}^{(0)}$ is calculated by imposing $w^{(0)}_{_{f(R)}}(z_{_{f(R)}}^{(0)})\,=\,0$ and by considering the Taylor expansion 

\begin{equation}
\frac{\sinh m\xi_0z}{m\xi_0z}\,\sim\,1+\frac{1}{6}(m\xi_0z)^2+\mathcal{O}(m\xi_0z)^4
\end{equation}
we obtain ${\displaystyle z_{_{f(R)}}^{(0)}\,=\,\frac{2\sqrt{6}}{\sqrt{3+(1+m\xi)e^{-m\xi}}}}$.
Since the stellar radius $\xi$ is given by definition $\xi\,=\,\xi_0\,z_{_{f(R)}}^{(0)}$, we obtain the  constraint

\begin{eqnarray}\label{radius_constraint}
\xi\,=\,\sqrt{\frac{3\Phi_c}{2\pi G}}\frac{1}{\sqrt{1+\frac{1+m\xi}{3}e^{-m\xi}}}
\end{eqnarray}
By solving numerically the constraint\footnote{In principle, there is a solution for any value of $m$.} Eq.(\ref{radius_constraint}), we find the modified expression of the radius $\xi$. If $m\,\rightarrow\,\infty$ we have the standard expression $\xi\,=\,\sqrt{\frac{3\Phi_c}{2\pi G}}$ valid for the Newtonian limit of GR. Besides, it is worth noticing  that in the  $f(R)$-gravity case, for $n=0$, the radius is smaller than in GR. On the other hand,  the gravitational potential $-\Phi$ gives rise to a deeper potential well than the corresponding Newtonian potential derived from GR \cite{stabile}. 

In the case $n\,=\,1$, Eq. (\ref{LEmod}) can be recast as follows

\begin{eqnarray}\label{LEmod2}
\frac{d^2\tilde{w}(z)}{dz^2}+\tilde{w}(z)\,=\,\frac{m\xi_0}{8}\int_0^{\xi/\xi_0}
dz'\,\biggl\{e^{-m\xi_0|z-z'|}-e^{-m\xi_0|z+z'|}\biggr\}\,\tilde{w}(z')
\end{eqnarray}
where $\tilde{w}\,=\,z\,w$. If we consider the solution of (\ref{LEmod2}) as a small perturbation to the one of GR, we have

\begin{eqnarray}\label{hypsol}
\tilde{w}^{(1)}_{_{f(R)}}(z)\,\sim\,\tilde{w}^{(1)}_{GR}(z)+e^{-m\xi}\Delta\tilde{w}^{(1)}_{_{f(R)}}(z)
\end{eqnarray}
The coefficient $e^{-m\xi}\,<\,1$ is the parameter with respect to which we perturb Eq. (\ref{LEmod2}). Besides these position  ensure us that when $m\,\rightarrow\,\infty$ the solution converge to something like 
$\tilde{w}^{(1)}_{GR}(z)$. Substituting Eq.(\ref{hypsol}) in Eq.(\ref{LEmod2}),  we have

\begin{eqnarray}
\frac{d^2\Delta\tilde{w}^{(1)}_{_{f(R)}}(z)}{dz^2}+\Delta\tilde{w}^{(1)}_{_{f(R)}}(z)\,=\,\frac{m\xi_0\,e^{m\xi}}{8}\int_0^{\xi/\xi_0}
dz'\,\biggl\{e^{-m\xi_0|z-z'|}-e^{-m\xi_0|z+z'|}\biggr\}\,\tilde{w}^{(1)}_{GR}(z')
\end{eqnarray}
and the solution is easily found

\begin{eqnarray}
w^{(1)}_{_{f(R)}}(z)\,\sim\,&&\frac{\sin z}{z}\biggl\{1+\frac{m^2{\xi_0}^2}{8(1+m^2{\xi_0}^2)}\biggl[1+\frac{2\,e^{-m\xi}}{1+m^2{\xi_0}^2}(\cos\xi/\xi_0+m\xi_0
\sin\xi/\xi_0)
\biggr]\biggr\}
\nonumber\\\nonumber\\&&
-\frac{m^2{\xi_0}^2}{8(1+m^2{\xi_0}^2)}\biggl[\frac{2\,e^{-m\xi}}{1+m^2{\xi_0}^2}(\cos\xi/\xi_0+m\xi_0\sin\xi/\xi_0)
\frac{\sinh m\xi_0z}{m\xi_0z}+\cos z\biggr]
\end{eqnarray}
Also in this case,  for $m\,\rightarrow\,\infty$, we do not  recover exactly $w^{(1)}_{GR}(z)$. The reason is the same of previous $n\,=\,0$ case \cite{eddington}. Analytical solutions for other values of $n$ are not available.

To conclude this section,  we report  the  gravitational potential profile generated by a spherically symmetric source of  uniform mass  with radius $\xi$.
We can impose a mass density of the form 
 \begin{equation}
 \rho\,=\,\frac{3M}{4\pi\xi^3}\Theta(\xi-|\mathbf{x}|)
 \end{equation}
  where $\Theta$ is the Heaviside function and $M$ is the mass \cite{stabile,cqg}. By solving field Eqs. (\ref{HOEQ}) {\it inside the star}  and considering the boundary conditions $w(0)\,=\,1$ and $w'(0)\,=\,0$, we get
\begin{eqnarray}\label{sol_pot}
w_{_{f(R)}}(z)\,=\,\biggl[\frac{3}{2\xi}+\frac{1}{m^2\xi^3}-\frac{e^{-m\xi}(1+m\,\xi)}{m^2\xi^3}\biggr]^{-1}
\biggl[\frac{3}{2\xi}+\frac{1}{m^2\xi^3}-\frac{{\xi_0}^2z^2}{2\xi^3}-\frac{e^{-m\xi}(1+m\,\xi)}{m^2\xi^3}\frac{\sinh m\xi_0z}{m\xi_0z}\biggr]
\end{eqnarray}
In the limit $m\,\rightarrow\,\infty$, we recover the GR case $w_{GR}(z)\,=\,1-\frac{{\xi_0}^2z^2}{3\xi^2}$. In Fig. \ref{fig} we show the behaviors of $w^{(0)}_{_{f(R)}}(z)$ and $w^{(1)}_{_{f(R)}}(z)$ with respect to the corresponding GR cases. Furthermore, we plot the potential generated by  a uniform spherically symmetric  mass distribution in GR and $f(R)$-gravity and the case $w^{(5)}_{GR}(z)$.

\section{Discussion and Conclusions}\label{conclu}

In this paper the hydrostatic equilibrium of a  stellar structure in the framework of $f(R)$-gravity has been considered. The study has been performed starting from the Newtonian limit of $f(R)$-field equations. Since the field equations satisfy in any case the Bianchi identity,  we can use the  conservation law of energy-momentum tensor. In particular adopting a polytropic equation of state relating the mass density to the  pressure, we derive the \emph{modified Lan\'{e}-Emden equation} and its solutions for $n\,=\,0,\,1$ which can be compared to the analogous solutions coming from the Newtonian limit of GR. When we consider the limit $f(R)\,\rightarrow\,R$, we obtain the standard hydrostatic equilibrium theory coming from  GR. A peculiarity of $f(R)$-gravity is the non-viability of Gauss theorem and then the \emph{modified Lan\'{e}-Emden equation} is an integro-differential equation where the mass distribution plays a crucial role. Furthermore the correlation between two points in the star is given by a Yukawa-like term of the corresponding  Green function. 

These solutions have been matched with those coming from   GR and the corresponding density radial  profiles have been derived. In the case $n\,=\,0$, we find an exact  solution, while, for $n\,=\,1$, we used a perturbative analysis with respect to the solution coming from GR. It is possible to demonstrate that   density radial profiles  coming   from $f(R)$-gravity analytic models and close to those coming from GR are compatible. This result   rules out  some wrong  claims in the literature stating that $f(R)$-gravity is not compatible with self-gravitating systems. Obviously the choice of the free parameter of the theory has to be consistent with boundary conditions and then   the solutions are parameterized by a suitable "wave-length" $m\,=\,\sqrt{-\frac{1}{3f''(0)}}$ that should be experimentally  fixed.

The next step is to derive self-consistent numerical solutions of \emph{modified Lan\'{e}-Emden equation} and build up  realistic   star models  where further   values of the polytropic index $n$ and other physical parameters, e.g.  temperature, opacity, transport of energy,  are considered. Interesting cases are the non-relativistic limit ($n\,=\,3/2$) and relativistic limit ($n\,=\,3$) of completely degenerate gas. These models  are a challenging task since, up to now,  there is no  self-consistent,  final   explanation for  compact objects (e.g. neutron stars) with masses larger than Volkoff mass, while observational evidences  widely indicate these objects \cite{mag}. In fact it is plausible that the gravity manifests itself on different characteristic lengths and  also  other contributions in the gravitational potential should be considered for these exotic objects. As we have seen above,  the gravitational potential well results modified by  higher-order corrections in the curvature. In particular, it is possible to show that  if we put in the action (\ref{FOGaction}) other curvature invariants also repulsive contributions can emerge \cite{stabile_2,cqg}. These situations have to be seriously taken into account in order to address several issues of relativistic astrophysics that seem to be out of the explanation range of the standard theory.

\section*{Acknowledgments}

This research was supported in part by INFN (Napoli-Trento, Italy)-MICINN 
(Barcelona, Spain) collaborative project.

\end{document}